# Correlation-distortion based identification of Linear-Nonlinear-Poisson models

Michael Krumin[1], Avner Shimron[1] and Shy Shoham[1*]

[1)] *Faculty of Biomedical Engineering, Technion IIT*
*32000 Haifa, Israel*

[*)] e-mail: sshoham@bm.technion.ac.il

Phone: +972-4-8294125

Fax: +972-4-8294599

**Abstract**

**Linear-Nonlinear-Poisson (LNP) models are a popular and powerful tool for describing encoding (stimulus-response) transformations by single sensory as well as motor neurons. Recently, there has been rising interest in the second- and higher-order correlation structure of neural spike trains, and how it may be related to specific encoding relationships. The distortion of signal correlations as they are transformed through particular LNP models is predictable and in some cases analytically tractable and invertible. Here, we propose that LNP encoding models can potentially be identified strictly from the correlation transformations they induce, and develop a computational method for identifying minimum-phase single-neuron temporal kernels under white and colored- random Gaussian excitation. Unlike reverse-correlation or maximum-likelihood, correlation-distortion based identification does not require the simultaneous observation of stimulus-response pairs – only their respective second order statistics. Although in principle filter kernels are not necessarily minimum-phase, and only their spectral amplitude can be uniquely determined from output correlations, we show that in practice this method provides excellent estimates of kernels from a range of parametric models of neural systems. We conclude by discussing how this approach could potentially enable neural models to be estimated from a much wider variety of experimental conditions and systems, and its limitations.**

*Keywords: system identification; correlation function; neural population; receptive field; point process; auto-regressive model*

## 1 Introduction

The characterization of neural encoding relationships is a central problem in experimental systems neuroscience. While general nonlinear systems characterization in terms of volterra-series expansions or detailed functional biophysical models have been applied to this problem, in practice, much of neural encoding research has been focused on reduced models like the Linear-Nonlinear-Poisson (LNP) cascade or its non-Poisson and mutually exciting generalizations like the Generalized Linear Models (Paninski et al. 2007). Encoding models are typically estimated using experiments where the neuron's response to a Gaussian stimulus process is measured: reverse-correlation is often applied to identify the LNP's linear kernel (Ringach and Shapley 2004; Schwartz et al. 2006; Wu et al. 2006), or maximum-likelihood-based optimization is used to identify the parameters of the more complex models.

In recent work (Krumin and Shoham 2009), we have obtained analytical results for the correlation structure of (multiple) LNP model outputs when the inputs are white Gaussian noise processes and the nonlinearities are either exponential, square or absolute-value. Using these results, we were able to generate synthetic spike trains with a fully-controllable correlation structure (defined by auto- and cross-correlation functions) by selecting appropriate linear kernels and nonlinearity parameters. This work was primarily motivated by our interest in neural pattern photo-stimulation (Shoham et al. 2005a; Farah et al. 2007), where synthetic spike trains can be used, for example, to emulate input activity onto a neuron, or to control neuron populations in artificial neuroprosthetic interfaces.

In this paper we extend our previous correlation-distortion results (relevant theory reviewed in section 2) and adapt them from the domain of synthetic spike train generation to the neural system identification problem. Viewed differently, this general approach can be used to determine parameters in a specific LNP encoding model excited by white Gaussian noise, given only the correlation structure of observed output spike trains. Unlike other neural system-identification methods, correlation-distortion based identification is not based on the joint statistics of stimulus-response pairs and thus does not require the simultaneous observation of stimulus and response. However, the method used in (Krumin and Shoham 2009) only considered uncorrelated white noise inputs and more-importantly calculates non-causal linear kernels which are implausible as neural encoding models. In section 3 we address this issue by using auto-regressive modeling of both input and output processes to develop a method for identifying plausible, minimum-phase, single-channel linear kernels, which we test in section 4. In section 5 we discuss the general features and possible applications of correlation-distortion based identification of neural encoding models and the fundamental and practical limitations of this approach. In what follows, we focus our discussion on the exponential nonlinearity, but the results can be easily extended to the other nonlinearities considered in (Krumin and Shoham 2009) and beyond.

## 2 Background - correlation distortions in a Linear-Exponential-Poisson model

In (Krumin and Shoham 2009) we calculated the auto- and cross-correlation functions of multiple-input multiple-output Linear-Nonlinear-Poisson systems excited by uncorrelated Gaussian noise. The uncorrelated processes are 'colored' by the linear stage and their correlation structure is systematically distorted by the static nonlinearity (Johnson 1994) in generating the rate processes. Interestingly, the expected auto- and cross-correlation functions of the Poisson spike trains have exactly the same structure as for the rates, except for the zero lag autocorrelation (proved in (Krumin and Shoham 2009)).

We will begin by briefly repeating the main derivation of the exponential correlation distortions. When transforming the Gaussian variables $X_i \sim N(0,1)$ with correlations $r_{ij} \triangleq E\left[X_i \cdot X_j\right]$ using the following transformation:

$$\Lambda_i = \exp\left(\mu_i + \sigma_i X_i\right) \qquad \textbf{(1)}$$

The resulting variable $\Lambda_i$ has a log-normal distribution with expectation:

$$E\left[\Lambda_i\right] = \exp\left(\mu_i\right) \cdot E\left[\exp\left(\sigma_i X_i\right)\right] = \exp\left(\mu_i\right) \cdot \exp\left(\frac{1}{2} E\left[\left(\sigma_i X_i\right)^2\right]\right) = \exp\left(\mu_i + \frac{\sigma_i^2}{2}\right) \qquad \textbf{(2)}$$

Where we have used Wick's theorem (Simon 1974) for mean zero, normal random variables $v \sim N\left(0, \sigma_v^2\right)$:

$$E\left[\exp\left(v\right)\right] = \exp\left(\frac{1}{2} E\left[v^2\right]\right) \qquad \textbf{(3)}$$

Using the same theorem, we can also derive expressions for the distorted correlations in terms of the correlations of Gaussian random variables:

$$\begin{aligned} R_{ij} &= E\left[\Lambda_i \cdot \Lambda_j\right] = E\left[\exp\left(\mu_i\right)\exp\left(\sigma_i X_i\right) \cdot \exp\left(\mu_j\right)\exp\left(\sigma_j X_j\right)\right] = \\ &= \exp\left(\mu_i + \mu_j\right) \cdot \exp\left(\frac{1}{2} E\left[\left(\sigma_i X_i + \sigma_j X_j\right)^2\right]\right) = E\left[\Lambda_i\right] E\left[\Lambda_j\right] \cdot \exp\left(\sigma_i \sigma_j r_{ij}\right) \end{aligned} \qquad \textbf{(4)}$$

And for the case $i = j$ ($r_{ii} = 1$):

$$R_{ii} \equiv E\left[\Lambda_i^2\right] = E^2\left[\Lambda_i\right] \cdot \exp\left(\sigma_i^2\right) \qquad \textbf{(5)}$$

Let us now assume that we observe the mean output rates and correlation structure, and want to estimate from them the pre-distorted statistical structure, and the parameters of the exponent. Equations (2) and (5) together allow us to evaluate the parameters $\mu_i$ and $\sigma_i$:

$$\sigma_i^2 = \ln\left(\frac{R_{ii}}{E^2\left[\Lambda_i\right]}\right) \qquad \mu_i = \ln\left(E\left[\Lambda_i\right]\right) - \frac{\sigma_i^2}{2} = \ln\left(\frac{E^2\left[\Lambda_i\right]}{\sqrt{R_{ii}}}\right) \qquad \textbf{(6)}$$

Finally, from equation (4) we derive the pre-distorted correlations as:

$$r_{ij} = \frac{1}{\sigma_i \sigma_j} \cdot \ln\left(\frac{R_{ij}}{E\left[\Lambda_i\right] E\left[\Lambda_j\right]}\right) \qquad \textbf{(7)}$$

## 3 Identifying a single Linear-Exponential-Poisson model

We consider the problem of identifying an LNP model with rate:

$$\lambda(t) = \exp\left(\mu + \sigma \cdot x(t)\right) = \exp\left(\mu + \sigma \cdot (h * s)(t)\right) \qquad \textbf{(8)}$$

from the measured auto-correlation function of the output spike train $R_{\Delta N}(\tau)$ and the auto-correlation function $R_s(\tau)$ of the Gaussian stimulus process $s(t)$. Here, $x(t)$ is the output after applying the linear kernel $h(t)$ to the stimulus ( $0 \le t \le K \cdot \Delta t$ ).

To solve this identification problem, we begin by using equation (6) to determine the parameters of the exponential nonlinearity, using the mean rate estimate: $E[\Lambda_i] = E[\Delta N]/\Delta t$ , and the rate's second moment estimate: $R_{ii} = E\left[\Delta N^2\right] - E[\Delta N]/\Delta t^2$ . Next, we observe that equation (7) relates between the correlation functions of $x(t)$: $R_x(\tau) \triangleq E\left[x(t+\tau)x(t)\right]$ and the rate auto-correlation function $R_\lambda(\tau) \triangleq E\left[\lambda(t+\tau)\lambda(t)\right]$, which in turn has the same expectation as the measurable auto-correlation functions $R_{\Delta N}(\tau)$ of the doubly stochastic Poisson process (for all $\tau \neq 0$ ).

### *3.1 Identifying the linear filter*

Given an estimate of the autocorrelation $R_x(\tau)$ of the linear stage output, the problem of identifying the filter $h(t)$ can be addressed using a number of frequency- or time-domain methods (Ljung 1999). The solution to this problem is fundamentally non-unique – since clearly, for example, isometric transformations like adding a delay to $h(t)$ or inverting the time axis will not alter the auto-correlation function. It is easy to see that in the frequency domain we have enough information to determine the correct magnitudes $\left|h(f)\right| = \sqrt{\mathcal{F}\left[R_x(\tau)\right]}$ but not the phases, leaving an infinite family of filters possible. In (Krumin and Shoham 2009) we estimated $h(t)$ using the matrix square root (Cholesky decomposition) of the auto-correlation matrix of $x(t)$, that is, a Toeplitz matrix which has $R_x(k\Delta\tau)$ along the *k*-th diagonal. Intuitively, we require that neural filters be causal, but this approach generally leads to non-causal filters. Another intuitive notion with regard to neural systems is that of stability with regard to perturbations (in general, neural systems can clearly be unstable – but those aspects probably cannot be studied using classical neural system identification experimental paradigms). Causal all-pole, auto-regressive (AR) filters that are also stable are also 'minimum phase', that is, they introduce minimal group delay to their input by having most of their response energy concentrated near the start of the impulse response. The minimum-phase constraint is thus a more plausible constraint for obtaining unique neural-like filters with the correct frequency transfer function, which can easily be estimated using auto-regressive modeling. We consider separately the cases of uncorrelated (white) and correlated (colored) stimulus processes (the process is schematically summarized in Figure 1).

The output of a discrete-time causal system driven by white Gaussian noise can be modeled using an auto-regressive (AR) model:

$$x[n] = -\sum_{k=1}^{p} a_k x[n-k] + s[n] \quad \textbf{(9)}$$

With the following z-domain transfer function:

$$H(z) = \frac{X(z)}{S(z)} = \frac{1}{1+\sum_{k=1}^{p} a_k z^{-k}} = \frac{1}{A(z)} \quad \textbf{(10)}$$

The parameters of the model ($a_k$) can be estimated directly from the auto-correlation function of the output $R_x(\tau)$ by solving the Yule-Walker equations (Makhoul 1975):

$$\begin{bmatrix} R_x(0) & R_x(1) & \cdots & R_x(p-1) \\ R_x(1) & R_x(0) & \cdots & R_x(p-2) \\ \vdots & \vdots & \ddots & \vdots \\ R_x(p-1) & R_x(p-2) & \cdots & R_x(0) \end{bmatrix} \cdot \begin{bmatrix} a_1 \\ \vdots \\ a_p \end{bmatrix} = -\begin{bmatrix} R_x(1) \\ \vdots \\ R_x(p) \end{bmatrix} \quad \textbf{(11)}$$

When the stimulus $s(t)$ is correlated, we can consider it as a white Gaussian noise process $w(t)$ that is filtered by a linear filter $S(z)$. The stimulus is subsequently filtered by the linear filter $H(z)$, resulting in another correlated Gaussian process $x(t)$ that enters the static non-linearity of the LNP model. To estimate the linear filter $h(t)$ from the auto-correlation function of the stimulus ($R_s(\tau)$) and the estimated auto-correlation of the linear stage output ($R_x(\tau)$), we model both $s(t)$ and $x(t)$ as auto-regressive processes with transfer functions $\frac{1}{A_S(z)}$ and $\frac{1}{A_X(z)}$ respectively, whose parameters are estimated separately from the auto-correlations. Dividing the latter by the former results in an auto-regressive-moving-average (ARMA) model that corresponds to the linear filter $h(t)$ which transforms between the two signals, with transfer function $H(z) = \frac{A_S(z)}{A_X(z)}$.

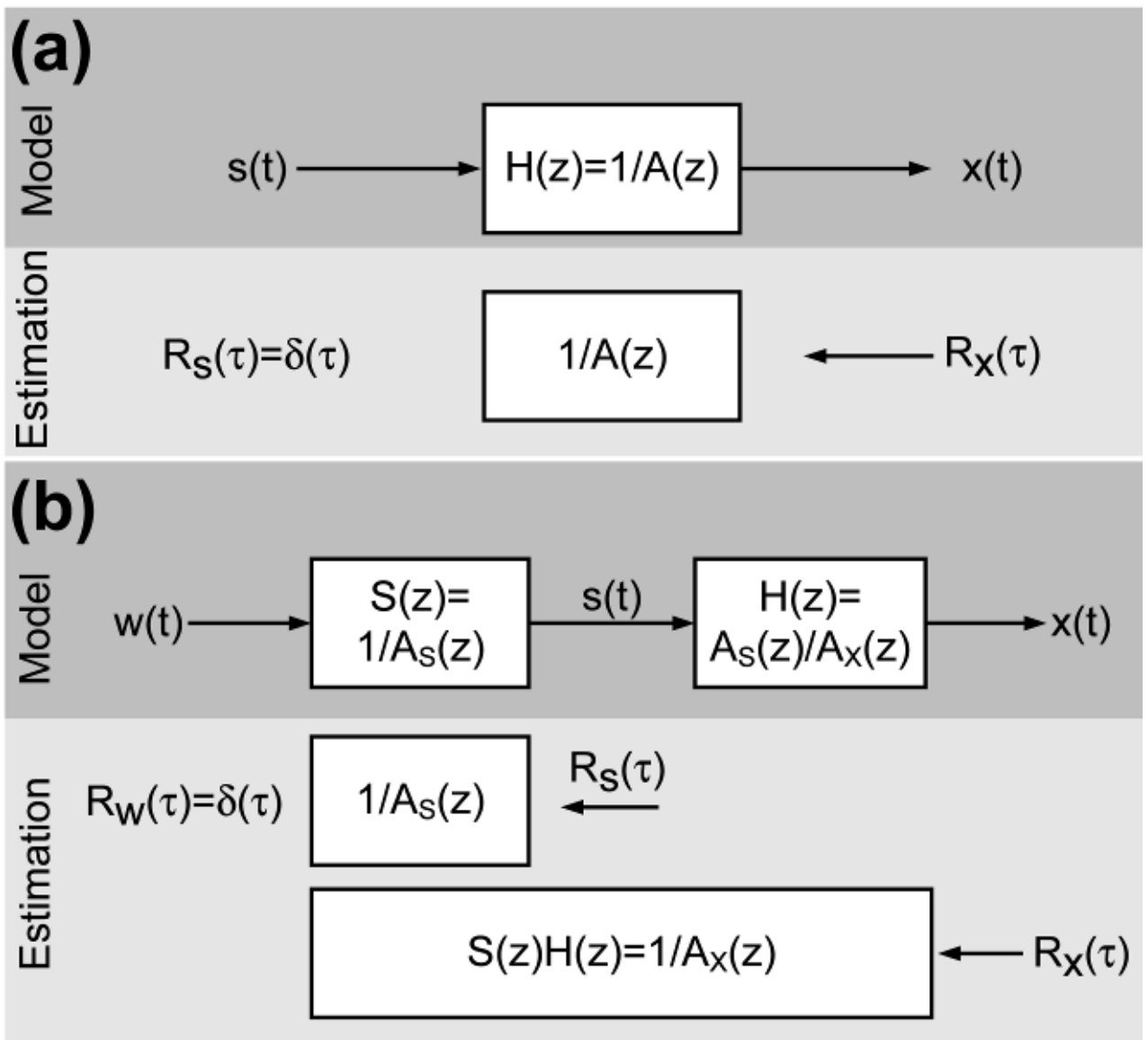

**Figure 1 AR and ARMA modeling of the linear stage. (a)** In the case of an uncorrelated stimulus s(t), the linear filter H(z) is approximated using an auto-regressive model 1/A(z), which is estimated from the output autocorrelation function $R_X(\tau)$. **(b)** In the case of a correlated stimulus s(t), the linear filter H(z) is approximated by an ARMA model $A_S(z)/A_X(z)$, which is estimated from the auto-correlation functions of the stimulus ($R_S(\tau)$) and the output ($R_X(\tau)$)

While these infinite-impulse-response AR or ARMA models fully capture the properties of the linear filter $h(t)$, we also transformed them into an equivalent finite impulse response (FIR) moving average (MA) representation, which is the familiar representation encountered in the field, by calculating the response to an impulse and truncating after the response decays.

### *3.2 Smoothing and stabilizing the solutions.*

Efficient estimation of the rate process' auto-correlation function $R_\lambda(\tau)$ is a crucial step in estimating the linear kernel. Spike trains' auto-correlation estimates tend to be very noisy and if directly used to estimate the rate autocorrelation lead to imprecise and unstable AR models. To stabilize the solution, we used the following steps:

**1.** A smoothing spline was applied with two constraints: a) $R_\lambda(0) = R_{\Delta N}(0) - E[\Delta N]/\Delta t^2$ and b) $R_\lambda(\tau = \tau_{\max}) = [E(\Delta N)/\Delta t]^2$ - the auto-correlation function has finite effective support and is constant outside.

**2.** The square system of Yule-Walker equations does not use the entire support of the auto-correlation function for a low order AR model, but for a high-order model it becomes ill-posed. We regularize the solution by lowering the order of the estimated auto-regressive model while solving an over-determined version of the Yule-Walker equations:

$$\begin{bmatrix} R_x(0) & R_x(1) & \cdots & R_x(\tilde{p}-1) \\ R_x(1) & R_x(0) & \cdots & R_x(\tilde{p}-2) \\ \vdots & \vdots & \ddots & \vdots \\ R_x(p-1) & R_x(p-2) & \cdots & R_x(p-\tilde{p}) \end{bmatrix} \cdot \begin{bmatrix} a_1 \\ \vdots \\ a_{\tilde{p}} \end{bmatrix} = - \begin{bmatrix} R_x(1) \\ \vdots \\ R_x(p) \end{bmatrix} \qquad \tilde{p} < p \qquad \textbf{(12)}$$

## 4 Results

We first tested these modeling and stabilization steps using simulations of an exponential LNP neuron model applied to white (Figure 2) and colored (Figure 3) zero-mean Gaussian-distributed pseudo-random stimuli processes. In both tests the sampling rate was 500Hz, the linear FIR kernel length was 25 samples, average firing rate was 20Hz, and the rate variance was 400 $Hz^2$. The record lengths used for the simulations were 1 and 10 hours respectively, and the AR/MA models used were of order 15 and 20, respectively. Our results demonstrate that in principle, when the neural kernel is exactly minimum-phase, this new approach allows accurate reconstruction of the linear kernel from autocorrelations, including the more noisy output spiking correlations (where the estimate appears noisier, at higher frequencies). The correlation-based estimator is generally noisier than the Spike Triggered Average, in part because it uses much less data (no stimulus-spike correlations). In the case of correlated-noise input we observed an estimator bias near $\tau=0$.

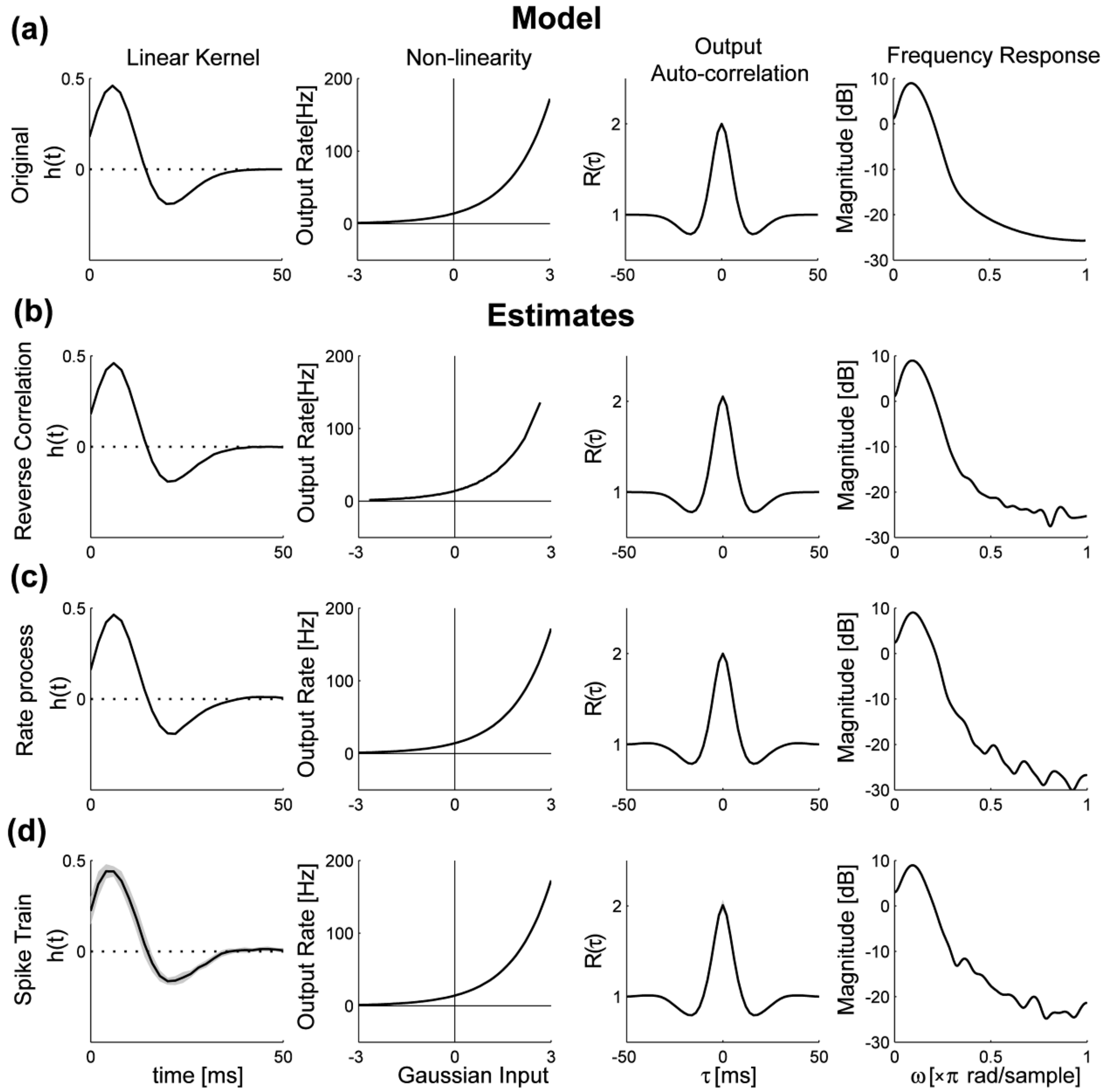

**Figure 2 Estimation of the linear filter kernel: white noise stimulus. (a)** Original filter, static non-linearity, auto-correlation function normalized by $E^2[\lambda]$ and filter frequency response. **(b)** Spike-Triggered-Average estimates. **(c)** Estimation from the (unobservable) auto-correlation function of the output rate process. **(d)** Estimation from the autocorrelation of the output spike train.

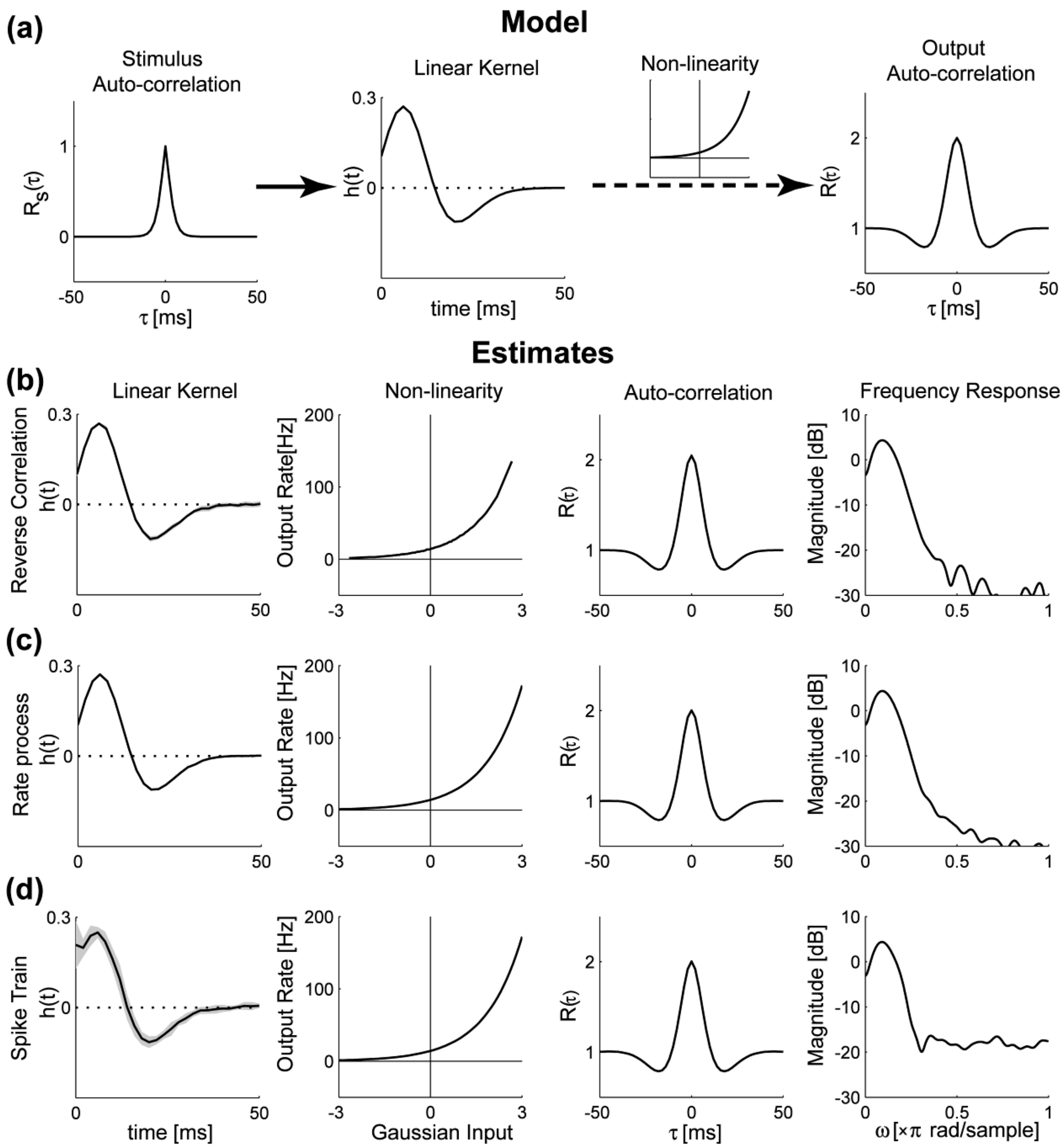

**Figure 3 Estimation of the linear filter kernel: colored noise stimulus. (a)** Stimulus auto-correlation passes through the linear filter and static non-linearity. Output auto-correlation function is normalized by $E^2[\lambda]$. **(b)** Spike-Triggered-Average estimates. **(c)** Estimation from the auto-correlation function of the output rate process. **(d)** Estimation from the autocorrelation of the output spike train (shadowed region represents the standard deviation of the estimator).

Next, we set out to test the method's applicability to estimating neural models. To address this issue, we generated a range of filter kernels from four parametric families of models previously used in the Computational Neuroscience literature to model temporal filter kernels. Model family #1 was used to model neurons in the retina and visual cortex in studies of contrast gain control (Yu et al. 2005; Yu and Sing Lee 2005):

$$h(t) = \sin(\pi\alpha t)\exp(-\beta t), \qquad t \geq 0 \tag{13}$$

Model family #2 was used to capture motion sensitivity in V1 neurons (Adelson and Bergen 1985; Dayan and Abbot 2001):

$$h(t) = \exp(\alpha t)\left(\frac{(\alpha t)^k}{k!} - \frac{(\alpha t)^{k+2}}{(n+2)!}\right), \qquad t \geq 0 \tag{14}$$

Where the parameter $k$ takes values of 3 or 5.

Model family #3 was used by Dayan and Abbott (2001) as a model of the temporal structure of retina and LGN neurons:

$$h(t) = \alpha^2 t \exp(-\alpha t) - \beta^2 t \exp(-\beta t), \qquad t \geq 0 \tag{15}$$

Finally, for completeness, we studied models from a fourth family that generalizes the alpha-function type kernels that are often used to model synaptic EPSPs (Dayan and Abbot 2001):

$$h(t) = (\alpha t)^k \exp(-\alpha t), \qquad t \geq 0 \tag{16}$$

We tested the method on $n = 99$ kernels from family 1 (parameter range: $10 \leq \alpha, \beta \leq 50\,\sec^{-1}$), $n = 102$ kernels from family 2 (parameter range: $k = 3,5$, $50 \leq \alpha \leq 100\,\sec^{-1}$), $n = 99$ kernels from family 3 (parameter range: $10 \leq \beta \leq 50\,\sec^{-1}$, $10 < \alpha \leq 250\,\sec^{-1}$, $\alpha > \beta$), and $n = 105$ kernels from family 4 (parameter range: $1 \leq k \leq 5$, $20 \leq \alpha \leq 200\,\sec^{-1}$). Two representative examples for each family of the original kernel, and the kernel estimated from the stochastic firing rate correlations are shown in Figure 4 (as the correlation-distortion based method cannot recover shift and scale, the estimated kernel was shifted and scaled to best match the original). Overall, the precise temporal structure of model filters was successfully estimated from the firing rate autocorrelation functions with correlation coefficients in the range of $\rho > 0.95$ (Figure 4).

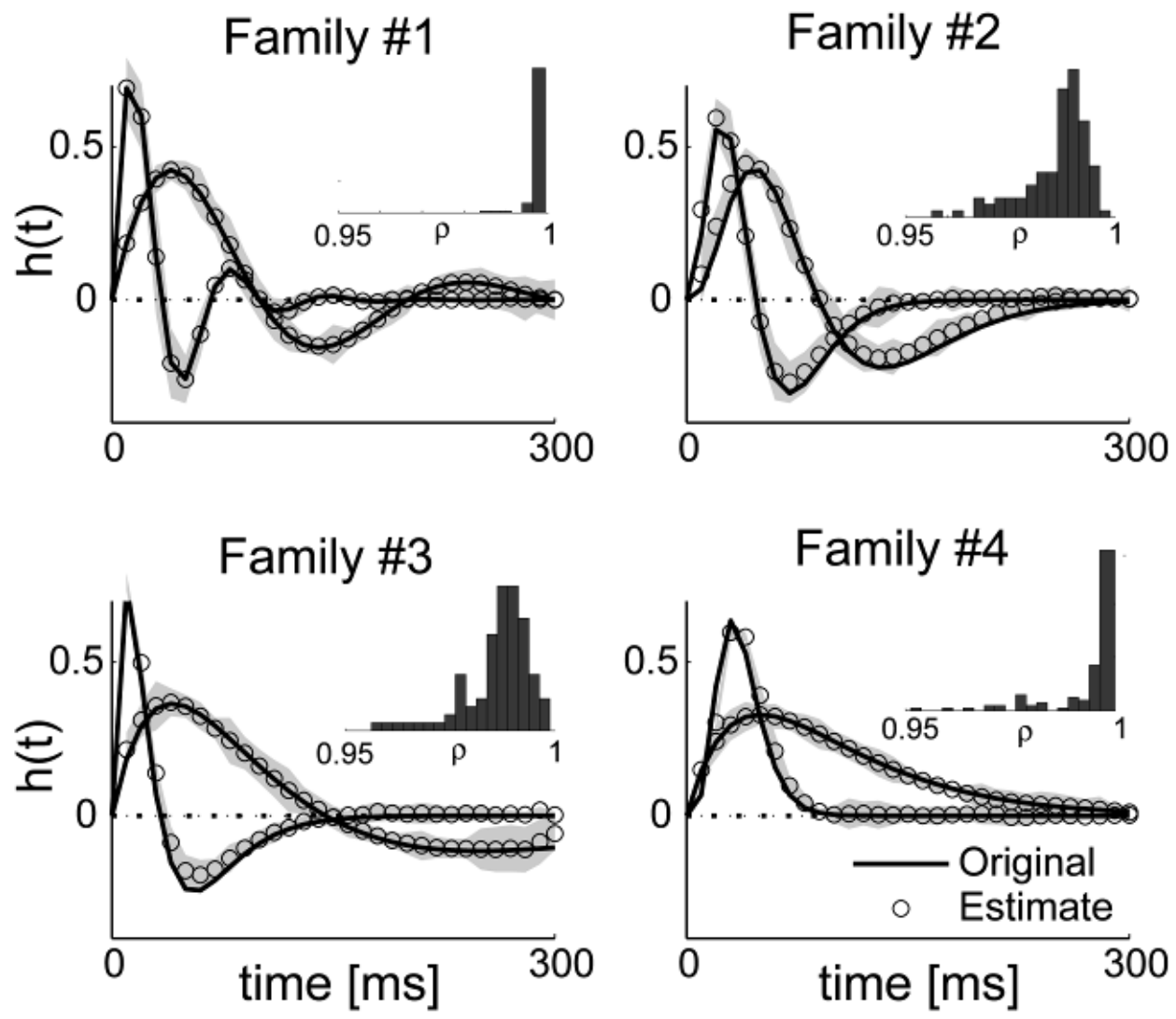

**Figure 4 Estimation of the linear filter kernel: examples.** Representative filters from four different parametric families (discussed in the text, equations (13)-(16)) are successfully estimated form the autocorrelation functions of the output and the stimulus. Note that the estimated filters are shifted in time and scaled in order to fit best to the original filters (shadowed region represents the standard deviation of the estimator). Insets: histograms of the correlation coefficient between the original and the estimated filters, separately to each parametric family.

To further understand the limitations of estimating neural models using this approach, we studied the worst-case estimate in our original data set (family #2, with $k = 5$, $\alpha = 50\ \text{sec}^{-1}$). As can be seen in Figure 5a, the original filter is not minimum-phase with some of its zeros residing outside the unit circle (Figure 5d). Applying the method to estimate the filter results in estimating its minimum-phase version. This minimum-phase filter has a very similar shape to the original filter and all its zeros lie inside the unit circle. Such similarity in shape between minimum-phase and not minimum-phase filters is typical for the four parametric families discussed above.

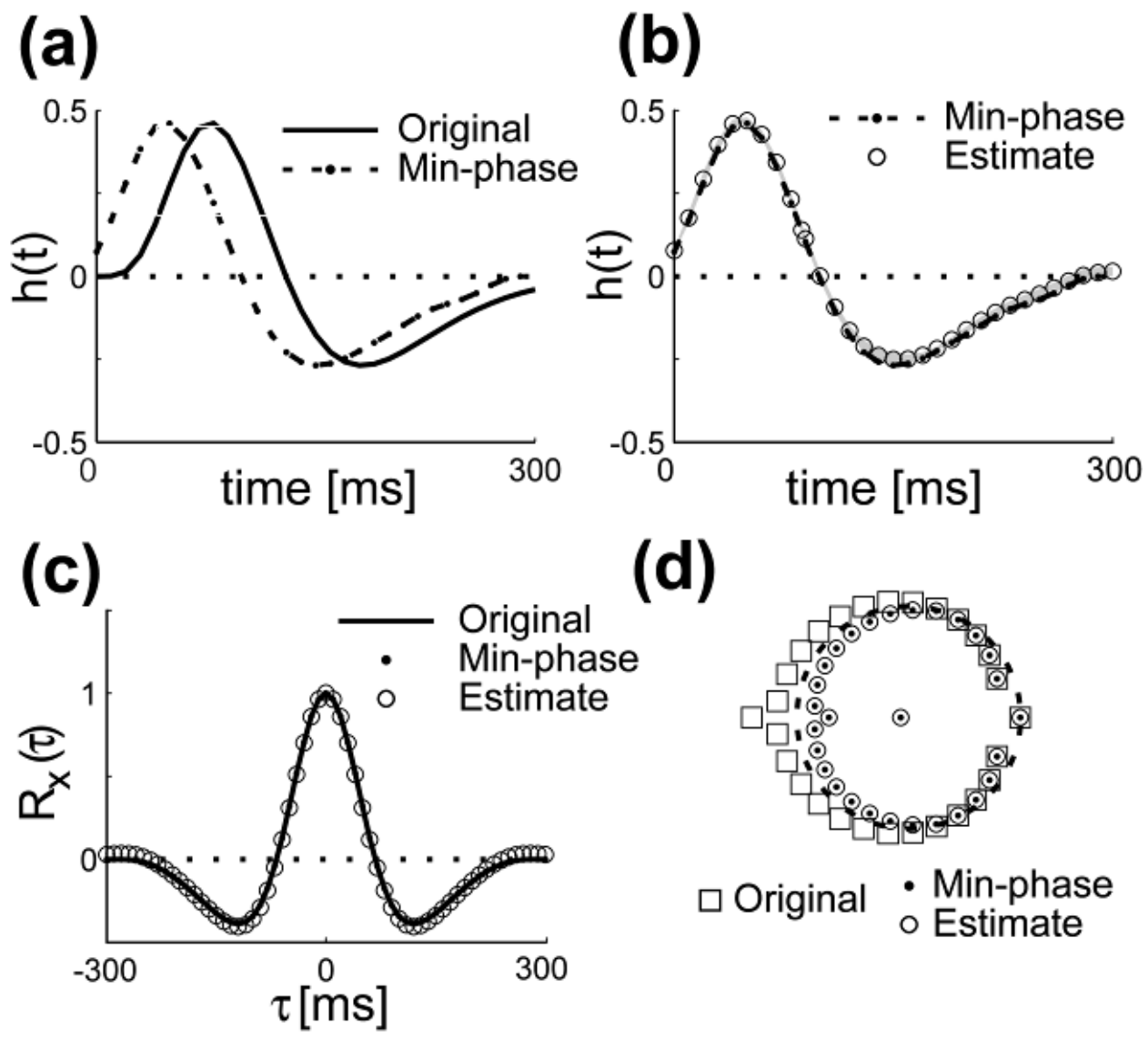

**Figure 5 Nonminimum-phase kernel estimation. (a)** Nonminimum-phase kernel and its minimum-phase form. Note that not only the delay is changed, but also the shape of the filter. **(b)** The estimated kernel fits the minimum-phase kernel. **(c)** The autocorrelation function is the same for all three linear kernels. **(d)** The zeros diagram for the three filters. Note that only for the original kernel part of its zeros lie outside the unit circle (dashed line), thus making it nonminimum-phase (one of the original filter zeros lies far on the left and is not presented here).

## 5 Discussion

In this paper we propose and develop a new method for indirectly identifying neuron encoding models from input-output correlation transformations by adapting our recent correlation-distortion results for LNP neuron models (Krumin and Shoham 2009). Our discussion focused on LNP encoding models with an exponential nonlinearity. Encoding relationships with exponential or exponential-like nonlinearities have been observed in the motor system (Paninski et al. 2004; Shoham et al. 2005b) and in various visual-system neurons (Gabbiani et al. 2002; Rust et al. 2006; Pillow et al. 2008), and may provide a good approximation in additional neuron models. Our development has been focused on a simple one-dimensional LNP neuron model and did not get into more complex self- and mutually-exciting encoding models (like the GLM). It is also instructive to mention related correlation distortion-type formulation by a number of authors exploring correlations induced in binary spike trains generated by directly thresholding Gaussian processes, without going through the intermediate step of a rate process (Johnson 1994; Dorn and Ringach 2003; Tchumatchenko et al. 2008; Macke et al. 2009). Because the second step is deterministic, the structure is single-stochastic, in difference with the doubly-stochastic Poisson structure explored here. Calculating the Gaussian-process correlations in this framework is somewhat more involved (requires numerical solutions of nonlinear equations). Experimental tests of our proposed method are currently underway.

To the best of our knowledge, this is a first example of a method proposed for blindly identifying a neuron encoding model, a task which is typically addressed using cross correlation-type analysis (Nykamp and Ringach 2002; Ringach 2004). This analysis requires precise measurement of each stimulus-response pair, which is challenging, for example, in cases where the subject is free to move and/or monitoring the exact stimulus to which the neural system is exposed to is prohibitively difficult. In this case the ability to recover encoding models only by independently measuring the input correlation structure and the response correlation structure, comprises a significant advantage over existing methods. Another possible scenario of interest is attempts to estimate the encoding function to internally generated, unobservable, stimuli like inputs from a lower level neural system, volitional movements in a paralyzed individual or dreams. Estimating the input correlations is an obvious (but not hopeless) challenge in these scenarios, as is correcting for the inevitable deviations from our underlying assumption of a stationary, Gaussian input process. Previous applications of a blind signal processing method in neurophysiological data analysis (namely, independent components analysis) addressed problems like the identification of functional groups within a large neuron population (Laubach et al. 1999) and spike sorting (Brown et al. 2001).

*Fundamental and practical limitations*

The main limitation that emerges in applying the correlation-distortion strategy is the non uniqueness of the kernel solutions with respect to changes that do not alter the output correlations, such as, for example, adding a time delay. Principally, the identified neural models are only unique and correct in terms of their frequency-magnitude response, and there is no one way of determining the phases from the correlation distortions. To determine a unique temporal filter with *plausible* causality, stability and time concentration properties (not necessarily *the actual* temporal kernel) we estimated (minimum-phase) AR or ARMA models, and then calculated their impulse response. While in general neural models are not necessarily minimum-phase (Victor 1989), our results suggest that this approach could under certain circumstances be effective at the estimation of a wide range of plausible neural models. Alternatively, one can estimate the same non-causal FIR filter kernels by calculating the matrix square root of the auto-correlation matrix (Krumin and Shoham 2009) and then find a minimum-phase FIR filter with the same magnitude frequency response by reflecting all the zeros into the unit circle. In future work, we will explore methods for removing some of the general indeterminacy in the estimation of the filter's phase structure using additional information from the higher-order correlation structure, multi-neural correlations or other sources.

Finally, we note that this approach is less general (requires more assumptions and choices) than the STA and/or ML estimators, including the assumption of a Gaussian input distribution, and having to select a specific non-linearity. In particular, a nonlinearity mismatch can result in a biased estimate, however, when we tested for the mismatch consequences we have seen fairly minor effects when the data was simulated using a non-exponential nonlinearity (Figure 6). Alternatively, such concerns can be addressed by using highly flexible nonlinearities or by comparing several different nonlinearities.

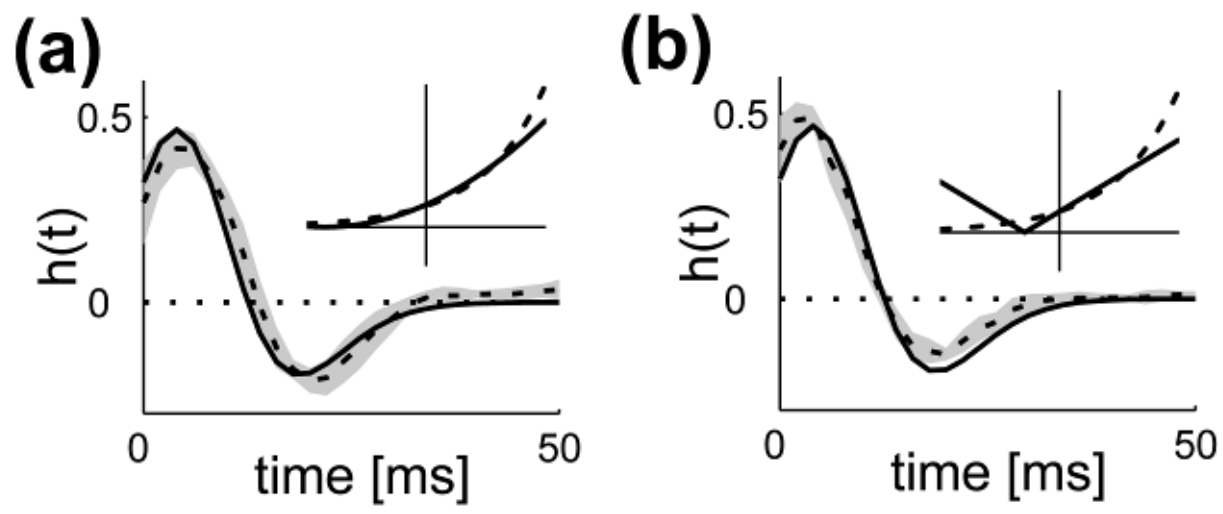

**Figure 6 Nonlinearity mismatch.** The original linear kernel (solid line) was estimated (dashed line) assuming exponential nonlinearity from the data that was simulated using **(a)** square or **(b)** absolute value nonlinearities respectively. Original nonlinearities (solid line) and estimated exponential nonlinearity (dashed line) are shown in the insets.

## Acknowledgements

This work was supported by Israeli Science Foundation grant #1248/06 and European Research Council starting grant #211055. We thank the two anonymous reviewers for their comments and suggestions.